\documentclass[11pt,dvips]{article}

\usepackage{epsfig}
\usepackage{graphicx,float}
\usepackage{array}
\usepackage{tabularx}
\usepackage{indentfirst}
\usepackage{amsmath,amssymb,bbold}
\usepackage{url}
\usepackage{rotating}
\usepackage{color}
\usepackage{hhline}
\usepackage{hyperref}
\usepackage{slashed}
\usepackage{tikz}
\usepackage{makecell}
\usepackage{pgfplots}
\usepackage{empheq}
\usepackage{physics}
\usepackage{cite}

\usepackage{ulem}

\newcommand{\s}{\\ \vspace*{-3.5mm}}

\renewcommand{\thefootnote}{\fnsymbol{footnote}}

\begin{document}


\begin{flushright}
\end{flushright}


\begin{center}
{\Large \bf Constructing the covariant three-point
            vertices systematically}\\[1.0cm]
{ Seong Youl Choi\footnote{sychoi@jbnu.ac.kr} and
  Jae Hoon Jeong\footnote{jaehoonjeong229@gmail.com}} \\[0.5cm]
{\it  Department of Physics and RIPC,
      Jeonbuk National University, Jeonju 54896, Korea}
\end{center}

\vskip 0.5cm

\begin{abstract}
\noindent
An algorithm is developed for efficiently constructing the Lorentz
covariant effective three-point vertices of the decay of a particle
into two daughter particles in which all the masses and spins of the
three particles can be arbitrary.
The closely-related one-to-one correspondence between the helicity
formalism and the covariant formulation is exploited for counting
the number of independent terms and identifying the basic covariant
three-point vertices. Assembling the basic operators according to
the developed algorithm enables us to construct all the
covariant three-point vertices systematically.
\end{abstract}



\vskip 0.8cm

\setcounter{footnote}{0}
\renewcommand{\thefootnote}{\alph{footnote}}

\section{Introduction}
\label{sec:introduction}

Despite the completion of the Standard Model (SM)~\cite{Glashow:1961tr}
of particle physics through the Higgs-boson discovery at the Large
Hadron Collider (LHC)~\cite{Aad:2012tfa},
various unsolved conceptual issues and unexplained experimental observations
suggest the SM to be an effective theory of a more fundamental theory.
One powerful strategy for probing new physics
beyond the SM (BSM) is to keep our studies as
{\it model-independent} as possible and to search for new particles
by including {\it spins higher than unity}.\s

Various high-spin (composite) hadrons have been discovered and
investigated~\cite{Zyla:2020zbs}.
A gravitino appears as a spin-3/2 supersymmetric partner of the massless
spin-2 graviton in supergravity~\cite{Nath:1975nj}.
The gravitational-wave discovery~\cite{Abbott:2016blz}
indicates the existence of massless spin-2 gravitons at the quantum level.
The massive spin-2 particles as the Kaluza-Klein (KK) excitations of the
massless graviton have been studied in extra-dimension
models~\cite{Antoniadis:1998ig}.
Recently, the scenario of high-spin dark matter (DM) particles has been
investigated
intensively~\cite{Babichev:2016hir}.
{\it For studying all these aspects, it is
crucial to probe all the allowed effective
interactions of particles of any spin as well as low-spin 
SM particles in a model-independent way.}  \s

In this work, we develop an efficient algorithm
for systematically constructing all the effective three-point vertices
consistent with {\it Lorentz invariance and locality}.
If necessary, other symmetry principles like local gauge invariance or
Bose/Fermi symmetries may be invoked.
Specifically, we consider the decay
of a massive particle of mass $m$ and spin $J$ into a massive
particle $M_1$ of mass $m_1$ and spin $s_1$ and a massive antiparticle 
$\bar{M}_2$ of mass $m_2$ and spin $s_2$,
while commenting how the massless case with $m_1=0$ or $m_2=0$ can
be accommodated straightforwardly. This study is a natural generalization
of two previous works having dealt with the identical-spin particles
of zero equal mass~\cite{Choi:2021ewa} and nonzero equal
mass~\cite{Choi:2021qsb}.\s

In this development, we adopt the conventional description of
integer and half-integer
wave tensors~\cite{Behrends:1957rup}
and utilize the closely-related equivalence between the helicity
formalism in the Jacob-Wick (JW) convention~\cite{Jacob:1959at}
and the standard covariant formulation. Their one-to-one
correspondence enables us to identify every basic building block
for {\it constructing the covariant three-point vertex corresponding to
every helicity combination explicitly}.\s

\setcounter{equation}{0}

\section{Characterization in the helicity formalism}
\label{sec:characterization_in_the_helicity_formalism}

\begin{figure}[htb]
\begin{center}
\includegraphics[width=8.3cm,height=5.cm]{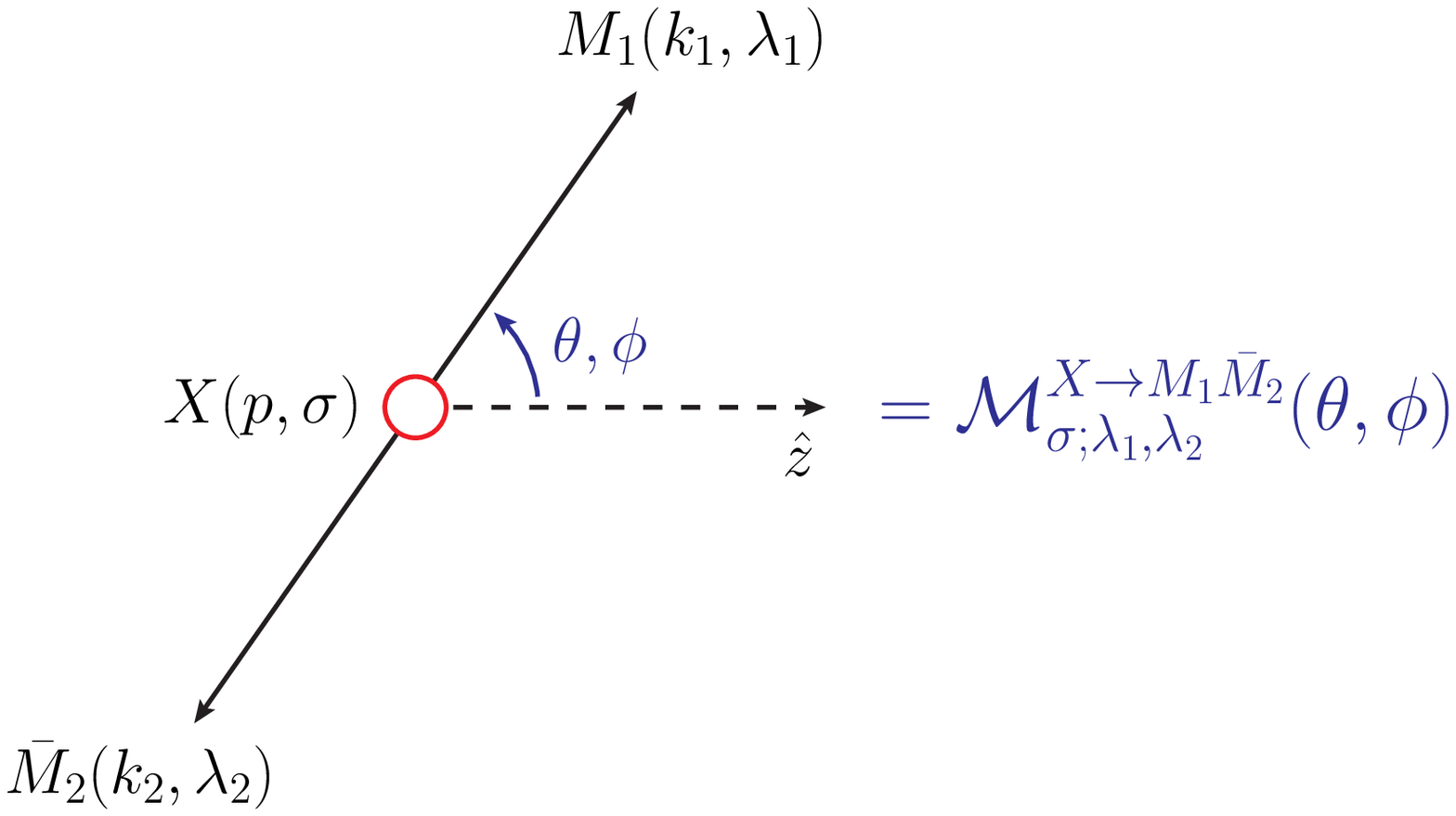}
\hskip 0.7cm
\includegraphics[width=8.8cm,height=5.cm]{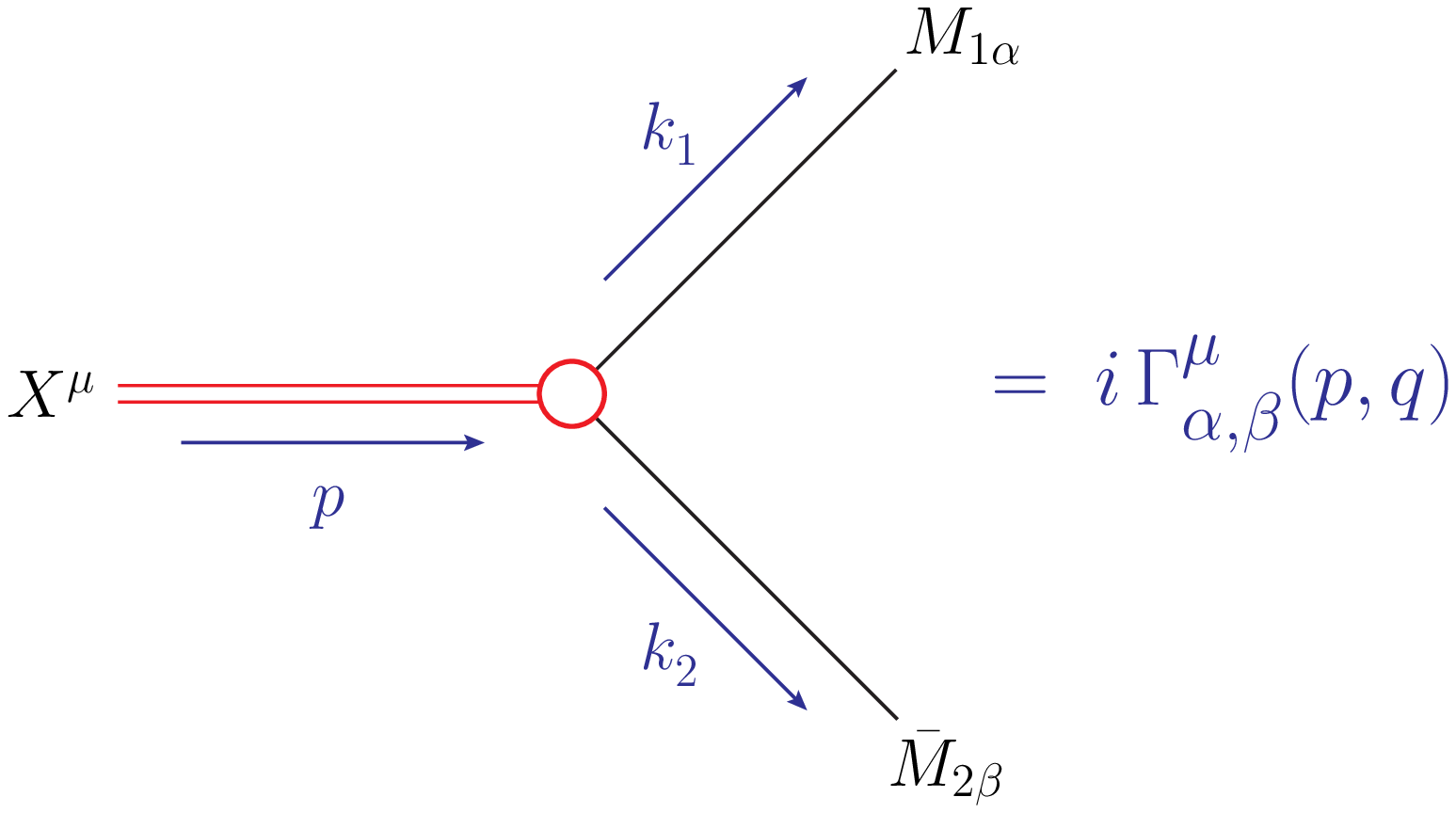}
\caption{\it (Left) Kinematic configuration for the
             helicity amplitude
             ${\cal M}^{X\to M_1 \bar{M}_2}_{\sigma;\,\lambda_1,\lambda_2}
             (\theta,\phi)$ of the two-body decay $X\to M_1\bar{M}_2$ of a
             massive particle $X$ into a particle $M_{1}$ and an antiparticle $\bar{M}_2$ in the $X$RF. The polar and
             azimuthal angles, $\theta$ and $\phi$, are defined with respect
             to an appropriately chosen coordinate system in the $X$RF.
             (Right) General $XM_1\bar{M}_2$ covariant three-point vertex $\Gamma^\mu_{\alpha,\beta}(p,q)$ for the decay of a particle $X$
             of spin $J$ and mass $m$ into a particle $M_{1}$ of 
             spin $s_1$ and mass $m_1$ and an antiparticle $\bar{M}_2$
             of spin $s_2$ and mass $m_2$. The indices, $\mu$, $\alpha$
             and $\beta$, stand for the sequences of $\mu=\mu_1\cdots \mu_{n}$,\,
             $\alpha_1\cdots \alpha_{n_1}$ and $\beta_1\cdots\beta_{n_2}$
             collectively with the non-negative integer $n=J$ or $n=J-1/2$ in the integer or half-integer spin $J$ case and the non-negative
             integers $n_{1,2}=s_{1,2}$ or $n_{1,2}=s_{1,2}-1/2$ in the
             integer or half-integer spin $s_{1,2}$ case, respectively.
             The symmetric and antisymmetric momentum combinations,
             $p=k_1+k_2$ and $q=k_1-k_2$, are introduced for the sake
             of notation.
}
\label{fig:kinematic_configuration_xm1m2_xrf}
\end{center}
\end{figure}

The helicity formalism~\cite{Jacob:1959at} allows us to
efficiently describe the two-body decay of a particle $X$ of spin $J$
and mass $m$ into a particle $M_{1}$ of spin $s_1$ and mass $m_1$ 
and an antiparticle $\bar{M}_2$  of spin $s_2$ and mass $m_2$. 
For the sake of a transparent analysis, we describe
the two-body decay $X\to M_1\bar{M}_2$ in the $X$ rest frame ($X$RF)
\begin{eqnarray}
X(p,\sigma)
   \ \ \rightarrow\ \
M_1(k_1,\lambda_1)\, +\, \bar{M}_2(k_2,\lambda_2)\,,
\label{eq:x_m1m2_decay}
\end{eqnarray}
with their momenta, $\{p, k_1, k_2\}$, and
helicities, $\{\sigma, \lambda_1, \lambda_2\}$, 
as depicted in the left side of Fig.~\ref{fig:kinematic_configuration_xm1m2_xrf}.\s

\vskip 0.5cm

The helicity amplitude of the decay $X\to M_1\bar{M}_2$ is decomposed in terms
of the polar and azimuthal angles, $\theta$ and $\phi$, defining the
momentum direction of the particle $M_1$ in a fixed coordinate system
as
\begin{eqnarray}
  {\cal M}^{X\to M_1\bar{M}_2}_{\sigma;\lambda_1,\lambda_2}(\theta,\phi)
\,\, =\,\, {\cal C}^J_{\lambda_1,\lambda_2} \,\,
  d^{J}_{\sigma,\,\lambda_1-\lambda_2}(\theta)\,
  e^{i(\sigma-\lambda_1+\lambda_2)\phi}
  \quad\ \ \mbox{with}\quad\ \
  |\lambda_1-\lambda_2|\leq J\,,
\label{eq:x_m1m2_helicity_amplitude}
\end{eqnarray}
in the JW convention~\cite{Jacob:1959at} (see the left side of
Fig.~\ref{fig:kinematic_configuration_xm1m2_xrf} for the kinematic 
configuration in the $X$RF), where the reduced helicity amplitudes ${\cal C}^J_{\lambda_1,\lambda_2}$ do not depend on the $X$ helicity
$\sigma$ due to rotational invariance and the polar-angle dependence is
fully encoded
in the Wigner $d$ function $d^{J}_{\sigma,\lambda_1-\lambda_1}(\theta)$ given
in the convention of Rose~\cite{merose2011}.
The helicity $\sigma$ of the spin-$J$ massive particle $X$ takes
one of $2J+1$ values between $-J$ and $J$. On the other hand, the helicities $\lambda_{1,2}$ of the spin-$s_{1,2}$ massive particles, $M_{1}$ and $\bar{M}_2$, can take one of $2s_{1,2}+1$ values between $-s_{1,2}$ and $s_{1,2}$, under
the constraint $|\lambda_1-\lambda_2|\leq J$. The number $n[J,s_1,s_2]$ of
independent reduced helicity amplitudes are given by
\begin{eqnarray}
  n\,[J,s_1,s_2]
\, =\,
  \left\{\begin{array}{ll}
           (2s_1+1)(2s_2+1)
         & \ \ \mbox{for}\ \ J\geq s_1+s_2\,, \\[2mm]
           (2s_1+1)(2s_2+1) -(s_1+s_2-J)(s_1+s_2-J+1)
         &  \ \ \mbox{for}\ \ |s_1-s_2| \leq J < s_1+s_2\,, \\[2mm]
           (s_1+s_2-|s_1-s_2|+1)\times(2J+1)
         & \ \ \mbox{for} \ \  J < |s_1-s_2|\,,
         \end{array}
  \right.
\label{eq:number_of_general_independent_terms}
\end{eqnarray}
that is valid for any allowed combinations of the spins $J$ and $s_{1,2}$.\s

{\it The master key to the algorithm for constructing the general three-point
vertices is to find every helicity-specific operator generating a reduced
helicity amplitude nonzero only for each specific helicity combination of $[\lambda_1,\lambda_2]$}. In the present work, we find  all the helicity-specific operators for the two-body decay $X\rightarrow M_1\bar{M}_2$ in the three cases, $(iii)$, $(ihh)$ and $(hhi)$, with the indices, $i$ and $h$, indicating whether the spins of $X$, $M_1$ and $\bar{M}_2$ are integer 
or half-integer, respectively.
Because the helicity-specific operators for the other processes such as 
those in the case $(hih)$ and for the charge-conjugated processes can 
be obtained by converting the derived ones according to simple 
symmetry arguments, as described 
explicitly in Sec.~\ref{subsec:conversion_of_the_helicity_specific_vertices}.\s

\setcounter{equation}{0}

\section{Bosonic and fermionic  wave tensors}
\label{sec:spin-j_spin-s_wave_tensors}

Generically, the decay amplitude of a particle $X$ of spin $J$ 
and mass $m$ into a particle $M_{1}$ of spin $s_1$ and mass $m_1$ and 
an antiparticle $\bar{M}_2$ of spin $s_2$ and mass $m_2$ can be written 
in terms of the covariant three-point vertex tensor $\Gamma$ 
(see the right side of Fig.~\ref{fig:kinematic_configuration_xm1m2_xrf}
for its diagrammatic description) as
\begin{eqnarray}
    {\cal M}^{X\to M_1\bar{M}_2}_{\sigma; \lambda_1,\lambda_2}
&=& \bar{\psi}_1^{\,\alpha_1\cdots\alpha_{n_1}}(k_1,\lambda_1)\,\,
    \Gamma^{\mu_1\cdots\mu_n}_{\alpha_1\cdots\alpha_{n_1},
           \beta_1\cdots\beta_{n_2}}(p,q)
     \,\, \psi_2^{\beta_1\cdots\beta_{n_2}}(k_2,\lambda_2)\,\,
    \psi_{\mu_1\cdots\mu_n}(p,\sigma)\,,
\label{eq:xm1m2_vertex}
\end{eqnarray}
with the non-negative integer $n=J$ or $n=J-1/2$ in the integer
or half-integer spin $J$ case and
the non-negative integers $n_{1,2}=s_{1,2}$ or $n_{1,2}=s_{1,2}-1/2$
in the integer or half-integer spin $s_{1,2}$ case, respectively.
$p$ and $\sigma$ are the momentum and helicity of the particle $X$,
and $k_{1,2}$ and  $\lambda_{1,2}$ are the momenta and helicities of two particles $M_1$ and $\bar{M}_2$, respectively. Here, the symmetric and antisymmetric
momentum combinations $p=k_1+k_2$ and $q=k_1-k_2$ of two momenta $k_1$ and $k_2$
are introduced.\s

We show the explicit expressions of the wave tensors of an on-shell
particle $X$ of spin $J$, mass $m$, momentum $p$ and helicity
$\sigma$~\cite{Behrends:1957rup}.
The wave tensors of the particles $M_1$ and $\bar{M}_2$ 
can be obtained by substituting $s_{1,2}$, $m_{1,2}$, $k_{1,2}$
and $\lambda_{1,2}$ for $J$, $m$, $p$ and $\sigma$  from the expressions given in
the following.\s

The bosonic particle $X$ of an integer spin $J=n$ with
a non-negative integer $n$ is represented by a rank-$n$ wave tensor
$\epsilon_{\mu_1\cdots\mu_{n}}(p,\sigma)$ as
\begin{eqnarray}
\psi_{\mu_1\cdots\mu_n}(p,\sigma)
\,=\, \epsilon_{\mu_1\cdots\mu_n}(p,\sigma)
\, =\,
   \sqrt{\frac{2^n\,(n+\sigma)!\, (n-\sigma)!}{(2n)!}}\,
   \sum^{1}_{\{\tau\}=-1}\,
   \delta_{\tau_1+\cdots+\tau_n,\,\sigma}\,
   \prod_{i=1}^n\,
   \frac{\epsilon_{\mu_i}(p,\tau_i)}{
         \sqrt{2}^{|\tau_i|}}\,,
\label{eq:explicit_bosonic_x_wave_tensors}
\end{eqnarray}
with the convention $\{\tau\}=\tau_1,\cdots,\tau_n$, which satisfies
the on-shell condition
$(p^2-m^2)\,\epsilon_{\mu_1\cdots\mu_n}(p,\sigma)=0$ for any helicity
value of $\sigma$ taking an integer value between $-n$ and $n$.
The bosonic wave tensor (\ref{eq:explicit_bosonic_x_wave_tensors})
satisfies
\begin{eqnarray}
      \varepsilon_{\alpha\beta\mu_i\mu_j}\,
      \epsilon^{\mu_1\cdots\mu_i\cdots\mu_j\cdots\mu_{n}}
       (p,\sigma)
\,=\, 0\,, \quad
      g_{\mu_i\mu_j}\,
      \epsilon^{\mu_1\cdots\mu_i\cdots\mu_j\cdots\mu_{n}}
      (p,\sigma)
\,=\, 0\,, \quad
      p_{\mu_i}\,
      \epsilon^{\mu_1\cdots\mu_i\cdots\mu_{n}}
      (p,\sigma)
\,=\, 0\,.
 \end{eqnarray}
as  it is {\it totally symmetric, traceless and divergence-free in the four-vector
indices}.
If the mass $m=0$, the wave tensor has only two
maximal-magnitude helicities $\sigma= \pm n=\pm J$ and its form is given simply
by a direct product of $n$ spin-1 wave vectors, each of which carries
the same helicity of $\pm 1$.\s

On the other hand, the fermionic particle $X$ of a half-integer spin $J=n+1/2$ with a non-integer value $n$ is represented in terms of the spin-1/2 $u$ and $v$ Dirac  spinors of a particle and an antiparticle. The fermionic particle and antiparticle wave tensors are given by~\cite{Behrends:1957rup}
\begin{eqnarray}
&& \mbox{ }\hskip -1.3cm
  \psi_{\mu}(p,\sigma)=u_{\mu_1\cdots \mu_n}(p,\sigma)
\, =\, \sum_{\tau=\pm 1/2}\,
   \sqrt{\frac{J+2\tau \sigma}{2J}}\,\,
   \epsilon_{\mu_1\cdots \mu_{n}}
    (p,\sigma-\tau)\,
   u(p,\tau)\,,
\label{eq:explicit_fermionic_wave_spinors_1}\\
&& \mbox{ }\hskip -1.3cm
  \psi_{\mu}(p,\sigma)=v_{\mu_1\cdots \mu_n}(p,\sigma)
\, =\, \sum_{\tau=\pm 1/2}\,
   \sqrt{\frac{J+2\tau \sigma}{2J}}\,\,
   \epsilon^*_{\mu_1\cdots\mu_{n}}
    (p,\sigma-\tau)\,
   v(p,\sigma)\,,
\label{eq:explicit_fermionic_wave_spinors_2}
\end{eqnarray}
where the spin-1/2 spinors satisfy their own on-shell conditions
$(\not\!{p}-m)\, u(p,\mbox{\small $\pm\frac{1}{2}$})=0$
and $(\not\!{p}+m)\, v(p,\mbox{\small $\pm\frac{1}{2}$})=0$. We note that
the massive wave spinors are {\it totally symmetric,  traceless and
divergence-free in the four-vector indices, as well}.
In the helicity amplitude (\ref{eq:xm1m2_vertex}), the fermionic $M_1$
wave tensor is given by $\bar{\psi}^\alpha_{1}(k_1,\lambda_1)
=\psi_{1}^{\alpha \dagger}(k_1,\lambda_1)\gamma^0$. If the mass $m=0$,
the particle or anti-particle $X$ wave tensor has
two maximal-magnitude helicities $\pm J$ and its form  is
given simply by a product of a $u$ or $v$ spinor and $n$ spin-1 wave vectors
with $n=J-1/2$.\s

We adopt the JW convention~\cite{Jacob:1959at} for deriving the spin-1 
vectors and spin-1/2 spinors in the $X$RF. We introduce three unit vectors
expressed in terms of the polar and azimuthal angles, $\theta$ and $\phi$ as
\begin{eqnarray}
   \hat{n}
= (\sin\theta\cos\phi,\, \sin\theta\sin\phi,\, \cos\theta)\,, \ \
   \hat{\theta}
= (\cos\theta\cos\phi,\, \cos\theta\sin\phi,\, -\sin\theta) \,, \ \
   \hat{\phi}
= (-\sin\phi,\, \cos\phi,\, 0)\,,
\label{eq:three_orthonormal_unit_vectors}
\end{eqnarray}
being mutually orthonormal, i.e.
$\hat{n}\cdot\hat{\theta}=\hat{\theta}\cdot\hat{\phi}
=\hat{\phi}\cdot\hat{n}=0$ and $\hat{n}\cdot\hat{n}=\hat{\theta}\cdot
\hat{\theta}=\hat{\phi}\cdot\hat{\phi}=1$.
In addition, we express the four-momentum sum $p=k_1+k_2$ and
the four-momentum difference $q=k_1-k_2$ as
\begin{eqnarray}
p\, =\,  m \hat{p} \quad \mbox{and} \quad
q\, =\, m (\omega_1^2-\omega_2^2)\,\hat{p}
     - m\kappa\, \hat{k}\,,
\label{eq:explicit_p_k_momenta_xrf}
\end{eqnarray}
in terms of two dimensionless re-scaled masses $\omega_{1,2}=m_{1,2}/m$
and two mutually orthogonal dimensionless four vectors $\hat{p}$ and $\hat{k}$, which are given by
\begin{eqnarray}
\hat{p}\,=\, (1, \vec{0}) \quad \mbox{and}\quad
\hat{k}\,=\, (0, \hat{n})\,.
\end{eqnarray}
in the $X$RF. The kinematic factor $\kappa=\eta^+\eta^-$ with
$\eta^\pm=\sqrt{1-(\omega_1\pm\omega_2)^2}$. These normalized momenta
$\hat{p}$ and $\hat{k}$, along with a few re-scaled mass-dependent
kinematic factors, can be exploited for expressing all the
reduced helicity amplitudes in the $X$RF.\s

The spin-1 wave vectors for the particle $X$ with momentum $p$ and two
particles, $M_{1}$ and $\bar{M}_2$, whose momenta
$k_{1,2}=(p\pm q)/2=m(e_{1,2},\,\pm\kappa \hat{n})/2$ with
$e_{1,2}=1 \pm (\omega^2_1-\omega^2_2)$, are given
in the JW convention by
\begin{eqnarray}
\epsilon(p,\pm 1) \! &=&\!
     \frac{1}{\sqrt{2}}\, (0,\, \mp 1,\,  -i,\, 0)\,, \ \ \ \ \ \ \
\epsilon(p,\, 0) \,\,\,\, =\,\,
     (0,\, 0,\, 0,\, 1)\,,
     \label{eq:xrf_x_wave_vectors} \\
\epsilon_1(k_1,\pm 1) \! &=&\!
     \frac{1}{\sqrt{2}}\, e^{\pm i \phi}\,
     (0,\, \mp \hat{\theta}-i \hat{\phi})\,, \ \ \
\epsilon_1(k_1, 0) \, =\,
     \frac{1}{2\,\omega_1}\,
     (\phantom{+}\kappa,\, e_1\hat{n})\,,
     \label{eq:xrf_m1_wave_vectors} \\
\epsilon_2(k_2,\pm 1) \! &=&\!
     \frac{1}{\sqrt{2}}\, e^{\mp i \phi}\,
     (0,\, \pm \hat{\theta}-i \hat{\phi})\,, \ \ \
\epsilon_2(k_2, 0) \, =\,
     \frac{1}{2\,\omega_2}\,
     (-\kappa,\, e_2\hat{n})\,,
     \label{eq:xrf_m2_0_wave_vector}
\end{eqnarray}
satisfying the relation
$\epsilon_2(k_2,\pm 1) = \epsilon_1(k_1, \mp 1)
=  -\epsilon^*_1(k_1,\pm 1) =  -\epsilon^*_2(k_2,\mp 1)$ in the $X$RF. \s

The spin-1/2 $u$ spinor of the particle fermion $X$
is given in the JW convention by
\begin{eqnarray}
&& u(p,\pm\mbox{\small $\frac{1}{2}$})
     \,=\, \sqrt{m}\,
      \left[\begin{array}{c}
            \xi_{\pm}(\hat z) \\[3mm]
            \xi_{\pm}(\hat z)
            \end{array}\right] \quad
             \mbox{with} \quad
             \xi_+(\hat z)=
     \left[\begin{array}{c}
            1 \\[3mm]
            0
            \end{array}\right] \;\;\mbox{ and }\;\;
             \xi_-(\hat z)=
     \left[\begin{array}{c}
            0 \\[3mm]
            1
            \end{array}\right]\,, \quad
\label{eq:spin-half_u_spinors}
\end{eqnarray}
in the $X$RF, and the spin-1/2 $u_{1}$ and $v_{2}$ spinors of the particle
and antiparticle fermions $M_1$ and $\bar{M}_{2}$ by
\begin{eqnarray}
&&  u_1(k_1,\pm\mbox{\small $\frac{1}{2}$})
     \,=\, \sqrt{\frac{m}{2}}\,
      \left[\begin{array}{c}
            \sqrt{e_1\mp\kappa}\, \chi_\pm(\hat{n}) \\[3mm]
            \sqrt{e_1\pm\kappa}\, \chi_\pm(\hat{n})
            \end{array}\right]\,, \quad
  v_2(k_2,\pm\mbox{\small $\frac{1}{2}$})
  \, = \, \sqrt{\frac{m}{2}}\,
      \left[\begin{array}{c}
            \pm\sqrt{e_2\pm\kappa}\, \chi_\pm(\hat{n}) \\[3mm]
            \mp\sqrt{e_2\mp\kappa}\, \chi_\pm(\hat{n})
                   \end{array}\right]
                   \,,
\label{eq:xrf_m_1_m_2_wave_spinors}
\end{eqnarray}
where the 2-component spinors $\chi_\pm(\hat{n})$ are written in the
$X$RF in terms of the angles, $\theta$ and $\phi$,
as
\begin{eqnarray}
\chi_+(\hat{n}) \,=\, \left[\begin{array}{l}
                       \cos\frac{\theta}{2}\, \\[2mm]
                       \sin\frac{\theta}{2}\, e^{i\phi}
                       \end{array}\right]\quad \mbox{and}\quad
\chi_-(\hat{n}) \,=\, \left[\begin{array}{c}
                       -\sin\frac{\theta}{2}\, e^{-i\phi} \\[2mm]
                       \!\!\!\!\!\!\cos\frac{\theta}{2} \,
                       \end{array}\right]\,,
\label{eq:2-component_spinor_wave_functions}
\end{eqnarray}
being mutually orthonormal, i.e. $\chi^\dagger_a(\hat{n})
\chi_b(\hat{n})=\delta_{a,b}$, with $a,b=\pm$ in the $X$RF.\s

\setcounter{equation}{0}

\section{Basic covariant three-point vertices}
\label{sec:basic_covariant_three_point_vertices}

In this section, we find all the Lorentz-covariant basic
bosonic and fermionic three-point operators by deriving the helicity-specific operators corresponding to the reduced helicity amplitudes for the
three spin combinations of $(J, s_1, s_2)=(1, 1, 1)$, $(1, 1/2, 1/2)$
and $(1/2, 1/2, 1)$ explicitly. The set of all these operators constitute
the backbone for constructing the covariant three-point vertices.\s

\subsection{Bosonic vertex operators}
\label{subsec:bosonic_raising_lowering_vertex_operators}

First, we consider the $1\to 1+1$ decay of a spin-1 particle $X$ into
two spin-1 massive vector bosons, $M_{1}$ and $\bar{M}_2$. The number of independent terms involving the $1\to 1+1$
decay is $n\,[1,1,1]=7$, accounting for the seven reduced helicity amplitudes,
${\cal C}^{\,1}_{0,0}$, ${\cal C}^{\,1}_{0,\pm 1}$, ${\cal C}^{\,1}_{\pm 1,0}$ and ${\cal C}^{1}_{\pm 1,\pm 1}$, in the $X$RF.
After a little manipulation, we find the five covariant
three-point vertex operators
\begin{eqnarray}
&&U^0_{\alpha\beta}\, \hat{k}_{\mu}
  \,\, =\,\, \hat{p}_{1\alpha }\hat{p}_{2\beta } \,\hat{k}_{\mu}
  \qquad\qquad\quad\quad \ \  \ \ \,\,\, \, \quad\leftrightarrow\qquad {\cal C}^{1}_{\,0,\,0} = \, \kappa^2\,,
  \label{eq:u0_bosonic_operators}\\[2pt]
&& U^\pm_{1\alpha\mu}\,\hat{p}_{2\beta }
  \,\, =\,\, \frac{1}{2}\, \left[g_{\bot\alpha\mu} \pm i \langle\alpha\mu\hat{p}\hat{k}\rangle\right]\hat{p}_{2\beta }
  \ \ \,\, \quad\leftrightarrow\qquad {\cal C}^{1}_{\pm 1,\,0} = \, \kappa\,,
  \label{eq:u1_bosonic_operators}\\
&& U^\pm_{2\beta\mu} \, \hat{p}_{1\alpha }
  \,\, =\,\, \frac{1}{2}\, \left[g_{\bot\beta\mu} \mp i \langle\beta\mu\hat{p}\hat{k}\rangle\right] \hat{p}_{1\alpha }
  \ \ \,\, \quad\leftrightarrow\qquad {\cal C}^{1}_{\,0,\pm 1} = \, -\kappa\,,
  \label{eq:u2_bosonic_operators}
\end{eqnarray}
and the two covariant composite operators $U^{\pm}$ of the contraction
of the basic operators $U^{\pm}_{1}$ and $U^{\pm}_{2}$ satisfying
\begin{eqnarray}
U^{\pm}_{\alpha\beta}\,\hat{k}_{\mu}
  \,\,\equiv\,\,
g^{\mu_1\mu_2}U_{1\alpha\mu_1}U_{2\beta\mu_2}\,\hat{k}_{\mu}
 \,\, = \,\,
\frac{1}{2}\, \left[\, g_{\bot\alpha\beta}
                  \pm i \langle\alpha\beta\hat{p}\hat{k}\rangle\,\right ]
                  \hat{k}_{\mu}
  \;\; \,\, \quad\leftrightarrow\qquad
                {\cal C}^{1}_{\pm 1,\pm 1} = \, -\kappa^2\,,
\label{eq:u_bosonic_operators}
\end{eqnarray}
expressed with two re-scaled momenta $\hat{p}_{1,2\alpha}=2\omega_{1,2}
\hat{p}_\alpha$ vanishing for $m_{1,2}=0$, the orthogonal tensor
$g_{\bot\mu\nu} =g_{\mu\nu}-\hat{p}_\mu\hat{p}_\nu
 +\hat{k}_\mu\hat{k}_\nu$ and $\langle \mu\nu\hat{p}\hat{k}\rangle
 =\varepsilon_{\mu\nu\rho\sigma}
\hat{p}^\rho\hat{k}^\sigma$ defined in terms of the totally antisymmetric
Levi-Civita tensor with the convention $\varepsilon_{0123}=+1$.
Each of the seven covariant
three-point vertices generates solely its corresponding reduced helicity
amplitude, as shown in Eqs.~\eqref{eq:u0_bosonic_operators}, \eqref{eq:u1_bosonic_operators}, \eqref{eq:u2_bosonic_operators}, and \eqref{eq:u_bosonic_operators}.\s

\vskip 0.5cm

\begin{figure}[htb]
\begin{center}
\includegraphics[width=5cm, height=5cm]{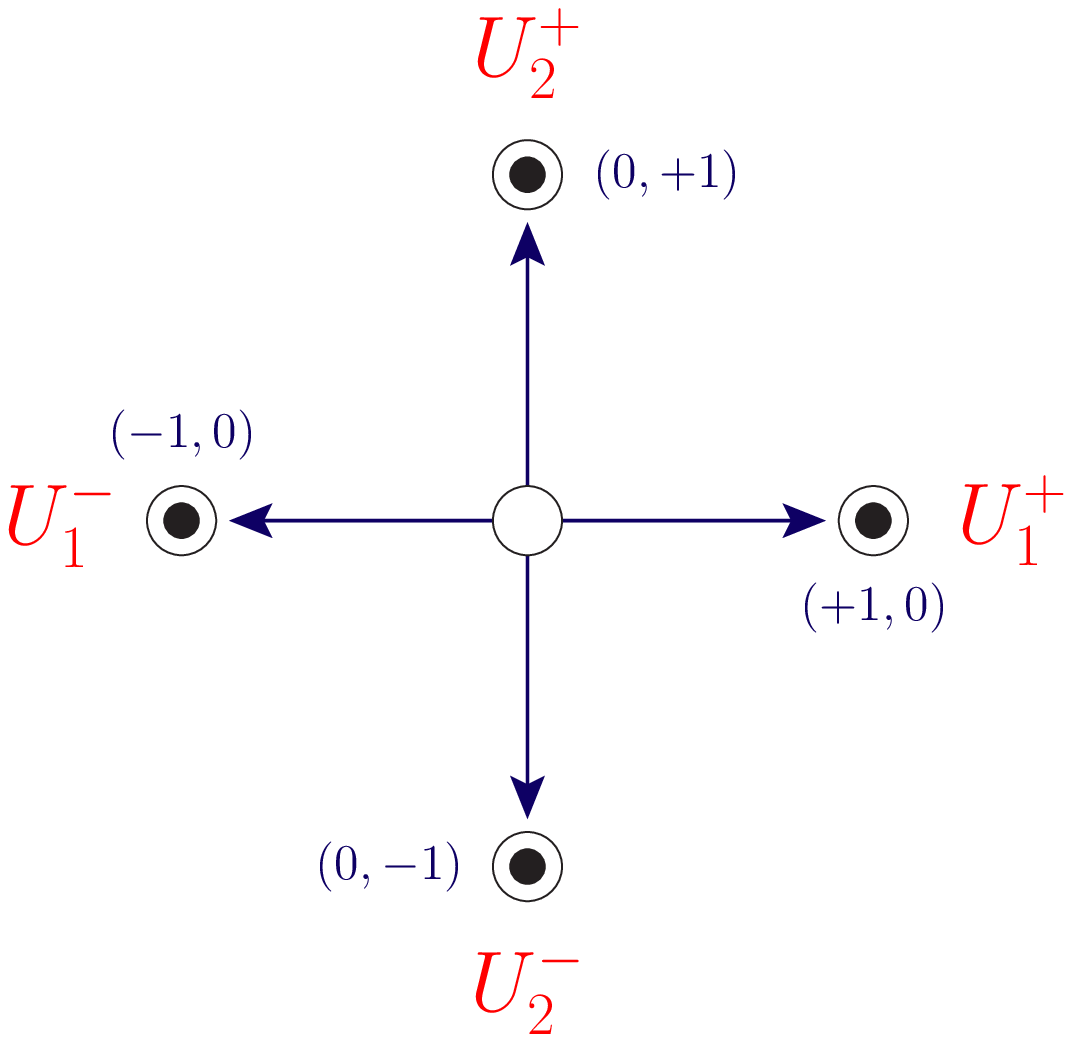}
\hskip 1.1cm
\includegraphics[width=5cm, height=5cm]{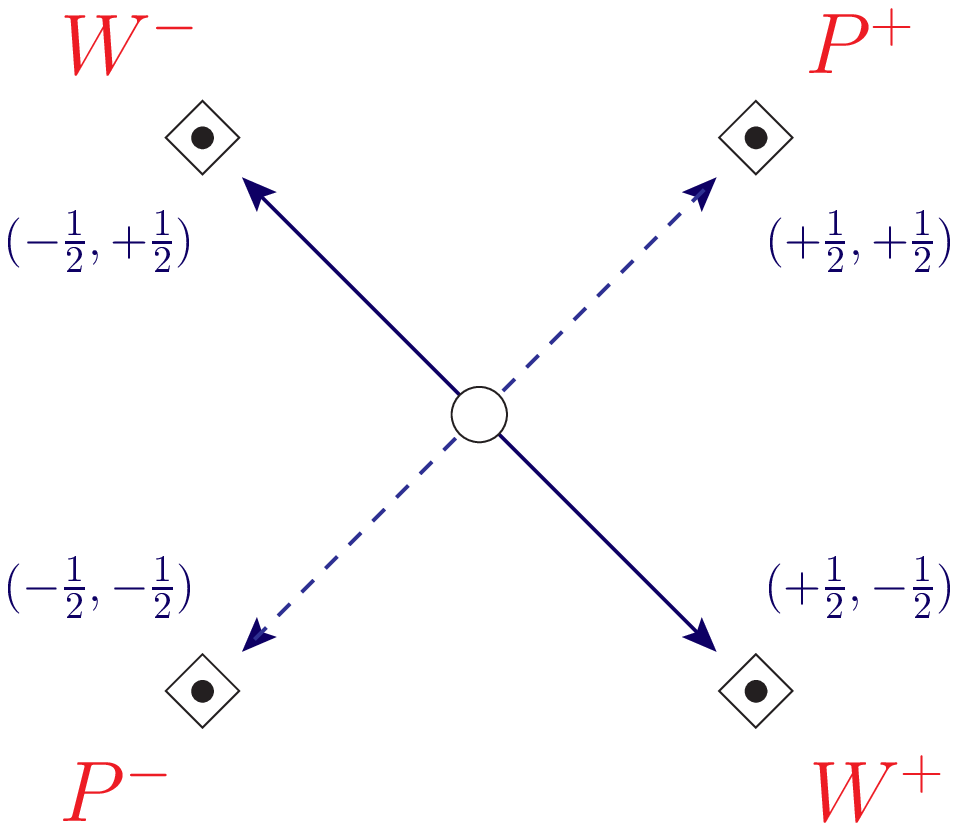}
\hskip 1.5cm
\includegraphics[width=3.8cm, height=5cm]{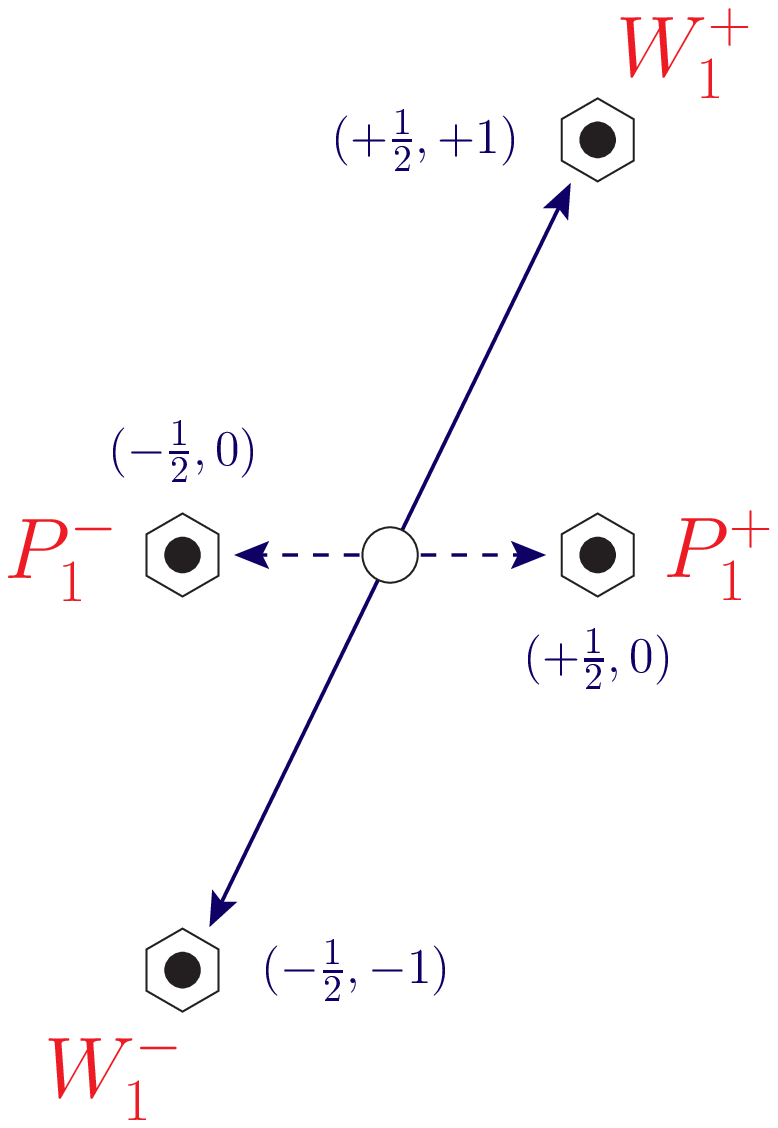}
\caption{\it Diagrammatic description of the four basic bosonic $U^\pm_1$
             and $U^\pm_2$ operators on the left-hand side,
             the four basic fermionic $P^\pm$ and $W^\pm$ operators
             in the middle, and the four basic fermionic operators
             $P^\pm_1$ and $W^\pm_1$ on the right-hand side.
             Their expressions are listed in Eqs.~\eqref{eq:u1_bosonic_operators}, \eqref{eq:u2_bosonic_operators}, \eqref{eq:p_fermionc_operators}, \eqref{eq:w_fermionc_operators}, \eqref{eq:p1_bosonic_fermionc_operators}, and \eqref{eq:w1_bosonic_fermionc_operators},
             respectively.
}
\label{fig:basic_operators}
\end{center}
\end{figure}

In order to clarify the essential role of each of the bosonic
basic operators, let us introduce an integer-helicity lattice
space consisting of $(2s_1+1)\times (2s_2+1)$ in order for each
point $[\lambda_1,\lambda_2]$ to stand for its corresponding
reduced helicity amplitude ${\cal C}^J_{\lambda_1,\lambda_2}$
existing only when $|\lambda_1-\lambda_2|\leq J$ and
$|\lambda_{1,2}|\leq s_{1,2}$. We can deduce that the one-step
horizontal and vertical transitions are dictated by the basic
operators, $U^{\pm}_1$ and $U^{\pm}_2$, from the point $[\lambda_1,\lambda_2]$
to the point $[\lambda_1\pm 1,\lambda_2]$ and the point
$[\lambda_1,\lambda_2\pm 1]$ in the helicity-lattice space, respectively, as shown in the left panel of Fig.~\ref{fig:basic_operators}.
The transitions by the basic operators, $U_1^{\pm}$ and $U_2^{\pm}$, enable us to deduce that the one-step diagonal transitions are dictated by the composite operators $U^{\pm}$ from the point $[\lambda_1,\lambda_2]$
to the point $[\lambda_1\pm 1,\lambda_2 \pm 1]$.
On the other hand, the three operators, $\hat{k}$, and $\hat{p}_{1,2}$, 
forming the operator $U_0$ and multiplied by the basic and composite 
operators generate no transition, that is to say, the helicity 
point remains intact by the three momentum operators,
solely changing the number of $\mu$, $\alpha$ and $\beta$
four-vector indices. Properly combining the six bosonic
raising and lowering operators, $U^\pm_1$, $U^\pm_2$, and $U^{\pm}$,
along with the three re-scaled momentum operators, $\hat{k}$
and $\hat{p}_{1,2}$, enables us to reach every integer-helicity
lattice point. {\it To summarize, for any
given integer $J$ and integer $s_{1,2}$, we can weave the covariant
three-point vertex corresponding to every integer-helicity combination of
$[\lambda_1,\lambda_2]$ efficiently and systematically.}
The explicit form of every covariant
three-point vertex consisting of the basic and composite operators with the re-scaled momentum operators for the integer $J$ and $s_{1,2}$ is to be presented in
Sec.~\ref{sec:weaving_covariant_three_point_vertices}. \s

\subsection{Fermionic vertex operators}
\label{subsec:fermionic_raising_lowering_vertex_operators}

First, we consider the decay of a spin-1 particle $X$ into
a spin-1/2 particle $M_1$ and a spin-1/2 antiparticle $\bar{M}_2$.
The number of independent terms involving the $1\to 1/2+1/2$ two-body
decay is $n\,[1,1/2,1/2]=4$, accounting for the four reduced
helicity amplitudes, ${\cal C}^{\,1}_{\pm 1/2,\pm 1/2}$ and
${\cal C}^{\,1}_{\pm 1/2, \mp 1/2}$. After a little manipulation,
we find the following four covariant three-point operators,
\begin{eqnarray}
&&  P^\pm \,\hat{k}_\mu =\, \frac{1}{2m}
                   (\eta^-\mp\eta^+ \gamma_5)\,
                   \hat{k}_\mu
            \quad\qquad\quad\ \; \,\leftrightarrow\qquad
   {\cal C}^{\,1}_{\pm 1/2, \pm 1/2} = \, -\kappa\,,
\label{eq:p_fermionc_operators} \\
&& W^\pm_\mu \, =\, \frac{1}{2\sqrt{2}m}\,
                   ( \eta^+\gamma^+_\mu
                      \pm \eta^-\gamma^-_\mu \gamma_5)
                   \qquad \ \ \, \leftrightarrow\qquad
   {\cal C}^{\,1}_{\pm 1/2, \mp 1/2} = \, \; \kappa\,,
\label{eq:w_fermionc_operators}
\end{eqnarray}
with the abbreviated kinematic parameters $\eta^\pm$ and the redefined
gamma matrices $\gamma^\pm_\mu$
\begin{eqnarray}
       \eta^\pm
\, =\, \sqrt{1-(\omega_1\pm\omega_2)^2} \quad \mbox{and}\quad
       \gamma^\pm_\mu
\, =\, \gamma_\mu + \frac{(\omega_1\pm \omega_2)\,\kappa}{
                          1-(\omega_1\pm\omega_2)^2}\, \hat{k}_\mu\,,
\label{eq:abbreviated_expressions_0}
\end{eqnarray}
satisfying the relation $\kappa=\eta^+\eta^-$. The corresponding reduced helicity amplitudes enable us to identify the basic fermionic operators, $P^\pm$ and $W^\pm$ with the normalized momentum $\hat{k}_{\mu}$ responsible for the one-half raising
and lowering diagonal and antidiagonal transitions in the half-integer
helicity lattice space  from the point $(0,0)$ to the points $(\pm 1/2,\pm 1/2)$
and $(\pm 1/2, \mp 1/2)$, respectively, as shown in the middle panel
of Fig.~\ref{fig:basic_operators}.\s

Second, we consider the decay $1/2\to 1/2+1$ of a spin-$1/2$ particle
fermion $X$ into a spin-$1/2$ particle fermion $M_1$ and a spin-$1$
antiparticle boson $\bar{M}_2$.
The number of independent terms involving the $1/2\to 1/2+1$ two-body
decay is $n\,[1/2,1/2,1]=4$, accounting for the four reduced
helicity amplitudes, ${\cal C}^{1/2}_{\pm 1/2,0}$ and
${\cal C}^{1/2}_{\pm 1/2, \pm 1}$. After a little manipulation,
we can find the following four covariant three-point operators,
\begin{eqnarray}
&& P^\pm_1\, \hat{p}_{2\beta}\, =\, \frac{1}{2m}\,
                   (\eta^-_1\mp\eta^+_1 \gamma_5)\,
                   \hat{p}_{2\beta}
            \qquad\qquad\,\, \leftrightarrow\qquad
   {\cal C}^{1/2}_{\pm 1/2, \, 0} = \, -\kappa^2\,,
\label{eq:p1_bosonic_fermionc_operators} \\
&& W^\pm_{1\beta} \, =\, \frac{1}{2\sqrt{2}m}\,
                   ( \eta^+_1\gamma^+_{1\beta}
                      \pm \eta^-_1\gamma^-_{1\beta} \gamma_5)
                   \qquad \ \ \; \leftrightarrow\qquad
   {\cal C}^{1/2}_{\pm 1/2, \pm 1} = \, - \kappa\,,
\label{eq:w1_bosonic_fermionc_operators}
\end{eqnarray}
with the abbreviated kinematic parameters and the redefined gamma
matrices $\gamma^\pm_{1\beta}$
\begin{eqnarray}
       \eta^\pm_1
\, =\, \sqrt{(1\pm\omega_1)^2-\omega^2_2} \quad \mbox{and}\quad
       \gamma^\pm_{1\beta}
\, =\, \gamma_\beta \mp \frac{2(1\pm \omega_1)}{
                          (1\pm\omega_1)^2-\omega^2_2}\, \hat{p}_\beta\,,
\label{eq:abbreviated_expressions_1}
\end{eqnarray}
satisfying the relation $\kappa=\eta^+_1\eta^-_1$. In this case, the two fermionic operators, $P^\pm_1$ and $W^\pm_1$ with the re-scaled momentum $\hat{p}_{2\beta}$, dictate the one-half step horizontal transitions from the point $(0,0)$ to the points $(\pm 1/2, 0)$,
and the one-half step horizontal and one-step vertical transitions from the point $(0,0)$
to the points $(\pm 1/2, \pm 1)$, respectively, as shown in the right panel of Fig.~\ref{fig:basic_operators},\s

Properly combining all the fermionic and bosonic operators with the three
re-scaled momentum operators enables us to reach every half-integer-helicity
lattice point. {\it To recapitulate, for any given $J$ and $s_{1,2}$, we can weave
the covariant effective three-point vertex corresponding to every helicity
combination of $[\lambda_1,\lambda_2]$ efficiently and systematically.}
The explicit form of the covariant three-point vertex constructed by
weaving the fermionic as well as bosonic operators is to be presented
in Sec.~\ref{sec:weaving_covariant_three_point_vertices}.\s

\setcounter{equation}{0}

\section{Weaving the covariant three-point vertices}
\label{sec:weaving_covariant_three_point_vertices}

Along with the re-scaled momenta, $\hat{p}_{1,2}$ and $\hat{k}$,
the four basic and two composite bosonic operators, $U^\pm_{1,2}$ and
$U^\pm$, and the two sets of basic fermionic operators,
$\{P^\pm, W^\pm\}$ and $\{P^\pm_1, W^\pm_1\}$, worked out
in Sec.~\ref{sec:basic_covariant_three_point_vertices}, enable us to
construct all the relevant covariant three-point vertices explicitly.
For this construction, it is crucial to take into account the feature
that bosonic and fermionic wave tensors are totally symmetric, traceless
and divergence-free in their four-vector indices and the fermionic
spinors satisfy
\begin{eqnarray}
    \gamma_{\alpha_i}
    u_1^{\alpha_1\cdots\alpha_i\cdots\alpha_{n_1}} (k_1,\lambda_1)
&=& \gamma_{\beta_i} v_2^{\beta_1\cdots\beta_i\cdots\beta_{n_2}}
    (k_2,\lambda_2)
\, =\,\, 0\,,
\label{eq:gamma_spinor_equation}
\end{eqnarray}
as well with the nonnegative integer $n_{1,2}=s_{1,2}-1/2$, so that every
fermionic vertex  involving $\gamma_{\alpha_i}$ or
$\gamma_{\beta_j}$ with $i=1,\cdots, n_1$ and $j=1,\cdots, n_2$
can be effectively excluded.
The $\mu$, $\alpha$ and $\beta$ four-vector indices in any
covariant three-point vertex can be  shuffled freely due to the totally
symmetric properties of the wave tensors, and any term including
$p_{\mu_i}$ for $i=1,\cdots n$ can be excluded effectively due to the
divergence-free condition. Moreover, the same condition allows us to
replace $k_{2\alpha_i}$ and $k_{1\beta_j}$ effectively
by $-p_{\alpha_i}$ and $p_{\beta_j}$ for $i=1,\cdots, n_1$ and
$j=1,\cdots, n_2$.\s

As many indices of different types are involved in expressing a
covariant three-point vertex especially for high-spin particles,
we introduce the following compact square-bracket notations
\begin{eqnarray}
&& [\,\hat{k}\,]^n
   \quad\ \ \rightarrow \quad
   (\hat{k}^n)_{\mu_1\cdots\mu_n}
   = \hat{k}_{\mu_1}\cdots\hat{k}_{\mu_n}\,,\\
&& [\,\hat{p}_1\,]^n
   \ \ \ \ \rightarrow \quad
   (\hat{p}_1^n)_{\alpha_1\cdots\alpha_n}
   = \hat{p}_{1\alpha_1}\cdots\hat{p}_{1\alpha_n}\,,\\
&& [\,\hat{p}_2\,]^n
   \ \ \ \ \rightarrow \quad
   (\hat{p}^n_2)_{\beta_1\cdots\beta_n}
   = \hat{p}_{2\beta_1}\cdots\hat{p}_{2\beta_n}\,,\\
&& [U^\pm_1]^n
   \quad\, \rightarrow  \quad
   (U^\pm_1)^n_{\alpha_1\cdots\alpha_n\mu_1\cdots\mu_n}
   = U^\pm_{1\alpha_1\mu_1}\cdots
     U^\pm_{1\alpha_n\mu_n}\,, \\
&& [U^\pm_2]^n
   \quad\, \rightarrow  \quad
   (U^\pm_2)^n_{\beta_1\cdots\beta_n\mu_1\cdots\mu_n}
   = U^\pm_{2\beta_1\mu_1}\cdots
     U^\pm_{2\beta_n\mu_n}\,, \\
&& [U^\pm]^n
  \quad\,\, \rightarrow \quad
   (U^\pm)^n_{\alpha_1\cdots\alpha_n\beta_1\cdots\beta_n}
   = U^\pm_{\alpha_1\beta_1}\cdots
     U^\pm_{\alpha_n\beta_n}\,,
\end{eqnarray}
for a non-negative integer $n$. Obviously, the zeroth power ($n=0$) of
any operator or re-scaled four momenta is set to be 1.
We emphasize once more that any
permutation of the $\alpha$, $\beta$ and $\mu$ four-vector indices
can be regarded to be equivalent as eventually the vertex operators
are to be coupled with the $X$ and $M_{1,2}$ wave tensors totally
symmetric in the four-vector indices. \s

\subsection{Bosonic and fermionic three-point vertices}
\label{subsec:bosonic_fermionic_three_point_vertices}

The helicity lattice point with any specific values of $\lambda_1$ and $\lambda_2$
can be reached by the helicity-specific operators $\mathcal{H}$ consisting of the three re-scaled momentum operators, $\hat{k}$ and $\hat{p}_{1,2}$ and the
helicity-specific transition operators $\mathcal{T}$ constructed with a product
of basic operators as
\begin{eqnarray}
       [\,{\cal H}^{J,s_1,s_2}_{A [\lambda_1,\lambda_2]}\,]
\, =\, [\,\hat{k}\,]^{J-|\lambda_1-\lambda_2|}\,
       [\,\hat{p}_1\,]^{s_1-|\lambda_1|}\,
       [\,\hat{p}_2\,]^{s_2-|\lambda_2|}\,
       [\,{\cal T}^{J,s_1,s_2}_{A [\lambda_1,\lambda_2]}\,]
       \quad \mbox{ with }\quad |\lambda_1-\lambda_2|\leq J
\label{eq:common_product_operator}
\end{eqnarray}
in an operator form where the index $A=iii$, $ihh$, and $hhi$ indicates whether the spins of $X$ and $M_{1,2}$ are integer $(i)$ or half-integer $(h)$, respectively.
For mathematical consistency, the powers of the re-scaled momentum operators, $\hat{p}_{1,2}$ and $\hat{k}$, should be non-negative and they play a crucial
role in determining the number of independent terms. \s

The helicity-specific operators in Eq.$\,$(\ref{eq:common_product_operator})
can be applied {\it even to the massless case with $m_1=0$ or $m_2=0$ simply 
by setting $[\,\hat{p}_1\,]^{s_1-|\lambda_1|}$ or
$[\,\hat{p}_2\,]^{s_2-|\lambda_2|}$ to unity}, because only
the maximal helicity values identical to the spin in magnitude
are allowed physically for a massless particle.\s

First, {\it in the $iii$ case with an integer $s_1$ and an integer $s_2$ forcing the spin $J$ to be an integer},
the transition vertices consisting of a sequence of the bosonic scalar 
and vector operators, $U^{\pm}$ and $U^{\pm}_{1,2}$, are 
classified by three regions as
\begin{eqnarray}
 [\,{\cal T}^{J,s_1,s_2}_{iii [\lambda_1,\lambda_2]}\,]
\, = \,
     \left\{\begin{array}{ll}
            [\, U^\pm\,]^{|\lambda_2|}\,
            [\, U_1^\pm\,]^{|\lambda_1-\lambda_2|}
          & \mbox{for} \ \ \lambda_{1,2}=\pm |\lambda_{1,2}| \ \
            \mbox{and} \ \ 0<|\lambda_2|\leq  |\lambda_1|\,, \\[2mm]
            [\, U^\pm\,]^{|\lambda_1|}\,
            [\, U_2^\pm\,]^{|\lambda_1-\lambda_2|}
          & \mbox{for} \ \ \lambda_{1,2}=\pm |\lambda_{1,2}| \ \
            \mbox{and} \ \ 0<|\lambda_1|< |\lambda_2|\,, \\[2mm]
            [\, U_1^\pm\,]^{|\lambda_1|}\,
            [\, U_2^\mp\,]^{|\lambda_2|}
          & \mbox{for} \ \ \lambda_1=\pm |\lambda_1| \ \
            \mbox{and} \ \ \lambda_2=\mp |\lambda_2|\,,
            \end{array}\right.
\label{eq:iii_helicity_specific_vertices}
\end{eqnarray}
in an operator form.\s

Second, {\it in the $ihh$ case with a half-integer $s_1$ and a half-integer
$s_2$ forcing the spin $J$ to be an integer}, the transition vertices consisting of the fermionic operators $P^\pm$ and $W^\pm$ as well as the bosonic scalar and vector operators, $U^{\pm}$ and $U^{\pm}_{1,2}$, are classified as
\begin{eqnarray}
 [\,{\cal T}^{J,s_1,s_2}_{ihh [\lambda_1,\lambda_2]}\,]
\, = \,
     \left\{\begin{array}{ll}
            [\,P^\pm\,]\, [\, U^\pm\,]^{|\lambda_2|-1/2}\,
            [\, U_1^\pm\,]^{|\lambda_1-\lambda_2|}
          & \mbox{for} \ \ \lambda_{1,2}=\pm |\lambda_{1,2}| \ \
            \mbox{and} \ \ |\lambda_2|\leq |\lambda_1|\,, \\[2mm]
            [\, P^\pm\,]\, [\, U^\pm\,]^{|\lambda_1|-1/2}\,
            [\, U_2^\pm\,]^{|\lambda_1-\lambda_2|}
          & \mbox{for} \ \ \lambda_{1,2}=\pm |\lambda_{1,2}| \ \
            \mbox{and} \ \ |\lambda_1|< |\lambda_2|\,, \\[2mm]
            [\, W^\pm\,]\, [\, U_1^\pm\,]^{|\lambda_1|-1/2}\,
            [\, U_2^\mp\,]^{|\lambda_2|-1/2}
          & \mbox{for} \ \ \lambda_1=\pm |\lambda_1| \ \
            \mbox{and} \ \ \lambda_2=\mp |\lambda_2|\,,
            \end{array}\right.
\label{eq:ihh_helicity_specific_vertices}
\end{eqnarray}
in an operator form.\s

Third, {\it in the $hhi$ case with a half-integer $s_1$ and an integer
$s_2$  forcing the spin $J$ to be a half-integer},  the transition vertices consisting of the fermionic operators $P_1^\pm$ and $W_1^\pm$ as well
as the bosonic scalar and vector operators, $U^{\pm}$ and $U^{\pm}_{1,2}$, are given by
\begin{eqnarray}
 [\,{\cal T}^{J,s_1,s_2}_{hhi [\lambda_1,\lambda_2]}\,]
\, = \,
     \left\{\begin{array}{ll}
            [\,P_1^\pm\,]\, [\, U^\pm\,]^{|\lambda_2|}\,
            [\, U_1^\pm\,]^{|\lambda_1-\lambda_2|-1/2}
          & \mbox{for} \, \, \lambda_{1,2}=\pm |\lambda_{1,2}| \, \,
            \mbox{and} \, \,\,1<|\lambda_2|< |\lambda_1|\,, \\[2mm]
            [\, W_1^\pm\,]\, [\, U^\pm\,]^{|\lambda_1|-1/2}\,
            [\, U_2^\pm\,]^{|\lambda_1-\lambda_2|-1/2}
          & \mbox{for} \, \, \lambda_{1,2}=\pm |\lambda_{1,2}| \, \,
            \mbox{and} \, \, \,|\lambda_1|< |\lambda_2|\,, \\[2mm]
            [\, P_1^\pm\,]\, [\, U_1^\pm\,]^{|\lambda_1|-1/2}\,
            [\, U_2^\mp\,]^{|\lambda_2|}
          & \mbox{for} \, \, \lambda_1=\pm |\lambda_1| \, \,
            \mbox{and} \, \, \lambda_2=\mp |\lambda_2|\,,
            \end{array}\right.
\label{eq:hhi_helicity_specific_vertices}
\end{eqnarray}
in an operator form.\s

{\it To conclude}, the general form of any covariant three-point vertex
$\Gamma_{A\alpha\beta;\mu}$ for any given $J$ and $s_{1,2}$
with $A=iii,\, ihh$, and $ hhi$ is a linear combination
of all the allowed helicity-specific three-point vertices.
The succinct operator form of the covariant three-point vertex is
given by
\begin{eqnarray}
      [\,\Gamma_A\,]
\,=\, \sum^{s_1}_{\lambda_1=-s_1}\,
      \sum^{s_2}_{\lambda_2=-s_2}\,
            c^{J,s_1,s_2}_{A [\lambda_1,\lambda_2]}\,
            [\, {\cal H}^{J,s_1, s_2}_{A [\lambda_1,\lambda_2]}\,]
            \quad \mbox{with} \quad
            A= iii,\, ihh,\, hhi\,,
\label{eq:general_covariant_three_point_vertices}
\end{eqnarray}
with the constraint $|\lambda_1-\lambda_2|\leq J$ where the
helicity-specific coefficients, $c^{J,s_1,s_2}_{A [\lambda_1,\lambda_2]}$
with $A=iii,\, ihh$, or $hhi$ depend only on the three masses, $m$ and
$m_{1,2}$. The expression (\ref{eq:general_covariant_three_point_vertices})
along with the helicity-specific covariant transition vertex operators
in Eqs.~(\ref{eq:iii_helicity_specific_vertices}),
(\ref{eq:ihh_helicity_specific_vertices}), and
(\ref{eq:hhi_helicity_specific_vertices}) is the key result of
the present work. Although it is originally deduced from the comparison
with the helicity amplitudes in the $X$RF, {\it the form is valid in
every reference frame because of its Lorentz-covariant form.} \s

\subsection{Conversion to all the other helicity-specific vertices}
\label{subsec:conversion_of_the_helicity_specific_vertices}

So far, we have derived the explicit forms of the helicity-specific operators
only for the spin combinations, $(iii)$, $(ihh)$, and $(hhi)$ in Sec.~\ref{subsec:bosonic_fermionic_three_point_vertices}.
However, the following simple symmetry arguments enable us to obtain the
covariant three-point vertices for all the remaining spin assignments.\s

Since any integer-spin wave tensor is given in the same form regardless of whether
the state is treated as a particle or an antiparticle, it is unnecessary to
consider the conversion of the helicity-specific operators
$ [\,{\cal H}^{J,s_1,s_2}_{iii [\lambda_1,\lambda_2]}\,]$.
In contrast, the helicity-specific operators
$[\,\bar{{\cal H}}^{J,s_1,s_2}_{ihh [\lambda_1,\lambda_2]}\,]$
for the decay of a integer spin-$J$ particle $X$ into an antiparticle $\bar{M}_1$
and a particle $M_2$ of half-integer spins $s_{1,2}$ can be given in an operator
form by
\begin{eqnarray}
     [\,\bar{{\cal H}}^{J,s_1,s_2}_{ihh [\lambda_1,\lambda_2]}\,]
\,=\,
    -[C \,{\cal H}^{J,s_1,s_2}_{ihh [\lambda_1,\lambda_2]}\,C^{-1}]\,.
\label{eq:conversion_rule_1}
\end{eqnarray}
This relation is derived by converting the wave spinors $\bar{u}_1^{\alpha}$
and $v_2^{\beta}$ of $M_1$ and $\bar{M}_2$ to  $v_1^{\alpha}$
and $\bar{u}_2^{\beta}$ of $\bar{M}_1$ and $M_2$ in the existing helicity-specific amplitudes by the unitary charge-conjugation $C$ giving the relation
$v=C\bar{u}^{ T}$. Similarily, the helicity-specific operators
$[\,\bar{{\cal H}}^{J,s_1,s_2}_{ihh [\lambda_1,\lambda_2]}\,]$ for the decay
of a half-integer spin-$J$ antiparticle $\bar{X}$ into a half-integer
spin-$s_1$ antiparticle $\bar{M}_1$ and a integer spin-$s_2$ particle
$M_2$ is obtained by the relation
\begin{align}
       [\,\bar{{\cal H}}^{J,s_1,s_2}_{hhi [\lambda_1,\lambda_2]}\,]
 \,=\,
     -[C \,{\cal H}^{J,s_1,s_2}_{hhi [\lambda_1,\lambda_2]}\,C^{-1}]\,,
\label{eq:conversion_rule_2}
\end{align}
in an operator form by converting the wave spinors $\bar{u}_1^{\alpha}$ and $u_{\mu}$ to $v_1^{\alpha}$ and $\bar{v}_{\mu}$ with the charge-conjugation
$C$. \s

On the other hand, the helicity-specific operators for the decay of a half-integer spin-$J$ particle $X$ of mass $m$ and helicity $\sigma$ into an integer spin-$s_2$ antiparticle $\bar{M}_1$ and a half-integer spin-$s_1$ particle $M_2$ of masses $m_{1,2}$ and helicities $\lambda_{2,1}$, respectively, are obtained simply
by the relation
\begin{eqnarray}
  ({\cal H}^{J,s_2,s_1}_{hih [\lambda_2,\lambda_1]})^{\mu_1\cdots
   \mu_n}_{\alpha_1\cdots \alpha_{n_2},\beta_1\cdots \beta_{n_1}}(p,q)
\,=\,
  ({\cal H}^{J,s_1,s_2}_{hhi [\lambda_1,\lambda_2]})^{\mu_1\cdots
   \mu_n}_{\beta_1\cdots \beta_{n_1},\alpha_1\cdots \alpha_{n_2}}(p,-q)\,,
\label{eq:conversion_rule_3}
\end{eqnarray}
through the replacements of $k_1 \leftrightarrow k_2$, $\alpha_1\cdots\alpha_{n_1} \rightarrow \beta_1\cdots\beta_{n_1}$, and $\beta_1\cdots\beta_{n_2} \rightarrow \alpha_1\cdots\alpha_{n_2}$.\s

\subsection{Off-shell electromagnetic gauge-invariant vertices}
\label{subsec:off-shell_EM_vertices}

Due to the electromagnetic (EM) gauge invariance, any off-shell
photon couples to a conserved current. Therefore, in any time-like
photon exchange process involving the $\gamma^* M_1 \bar{M}_2$ vertex,
the off-shell photon can be treated as a spin-1 particle of mass
$m=\sqrt{p^2}$. Moreover, the covariant three-point
$\gamma^* M_1\bar{M}_2$ vertex can be cast into a manifestly
EM gauge-invariant form~\cite{Boudjema:1990st} as
\begin{eqnarray}
     \Gamma^\mu_{X{\rm EM}\,\,
      \alpha,\beta}
\, =\,
     p^2\, \Gamma^\mu_{\alpha,\beta}
     - (\, p\cdot \Gamma_{\alpha,\beta}\,)\, p^\mu\,,
\label{eq:gauge_invariant_photon_gamma_m1_m2_three_point_vertex}
\end{eqnarray}
satisfying the current conservation condition
$p_\mu\Gamma^\mu_{X{\rm EM}\, \alpha,\beta}=
p\cdot\Gamma_{X{\rm EM}\, \alpha,\beta}=0$ automatically. \s

Similarily, in any process involving the $X \gamma^*\bar{M}_2$ vertex,
the off-shell photon $\gamma^*$ can be treated as a spin-1 particle
of mass $m_1=\sqrt{k^2_1}$. Then, the covariant three-point vertex can
be cast into a manifestly EM gauge-invariant form as
\begin{eqnarray}
     \Gamma^\mu_{1{\rm EM}\,\,\alpha,\beta}
\, =\,
     k_1^2\, \Gamma^\mu_{\alpha,\beta}
     - k_1^\rho\, k_{1\alpha}\, \Gamma^\mu_{\rho,\beta}\,,
\label{eq:gauge_invariant_photon_m1_gamma_m2_three_point_vertex}
\end{eqnarray}
satisfying the current conservation condition
$k_1^\alpha \, \Gamma^\mu_{1{\rm EM}\, \alpha,\beta}=0$ automatically. \s

\setcounter{equation}{0}

\section{Conclusions}
\label{sec:conclusions}

We have developed an efficient algorithm for constructing all the covariant
effective three-point vertices for the decay of a particle $X$ of
spin $J$ and mass $m$ into a particle $M_{1}$ and an antiparticle 
$\bar{M}_2$ with any spins and masses, $s_{1,2}$ and $m_{1,2}$. 
For this development, we have exploited the closely-related equivalence 
between the helicity formalism and the covariant formulation for identifying 
the basic operators and then for constructing all the covariant three-point
vertices.\s

We have presented all the helicity-specific covariant three-point vertices
in an operator form in Eqs.~(\ref{eq:iii_helicity_specific_vertices}),
(\ref{eq:ihh_helicity_specific_vertices}) and
(\ref{eq:hhi_helicity_specific_vertices}) explicitly in the $iii$, $ihh$
and $hhi$ cases, respectively. We have listed the conversion rules to all the
other cases in Eqs.$\,$(\ref{eq:conversion_rule_1}), (\ref{eq:conversion_rule_2})
and (\ref{eq:conversion_rule_3}). In addition, we have shown that the case with
$m_1=0$ or $m_2=0$ can be accommodated straightforwardly. Finally, we
have described how to obtain the EM gauge-invariant vertices involving
a virtual photon in the initial or final state.\s

The general algorithm for constructing the covariant three-point vertices 
enables us to  work out various theoretical and phenomenological aspects
systematically and efficiently, including the indirect and direct searches 
for DM particles of any spin and the production of new particles of any 
spin at high energy colliders. An interesting issue to be pursued is whether 
the bosonic and fermionic cases can be synthesized further.\s

\section*{Acknowledgments}
\label{sec:acknowledgments}

The work was in part by the Basic Science Research Program of Ministry of
Education through National Research Foundation of Korea
(Grant No. NRF-2016R1D1A3B01010529) and in part by the CERN-Korea theory
collaboration.\s

\end{document}